\documentclass[12pt]{article}
\usepackage{amsmath}
\usepackage{graphicx, epsfig}

\begin{document}

\title{A tomographic analysis of reflectometry data I: Component
factorization}
\author{{\small Fran\c{c}oise Briolle}\thanks{%
Corresponding author, francoise.briolle@univmed.fr} \thanks{%
Centre de Physique Th\'{e}orique, CNRS Luminy, case 907, F-13288 Marseille
Cedex 9, France}, {\small Ricardo Lima}\footnotemark[2] , {\small Vladimir
I. Man'ko}\thanks{%
P. N. Lebedev Physical Institute, Leninskii Prospect 53, Moscow 117924,
Russia, e-mail: manko@sci.lebedev.ru} {\small and R. Vilela Mendes}\thanks{%
IPFN - EURATOM/IST Association, Instituto Superior T\'{e}cnico, Av. Rovisco
Pais 1, 1049-001 Lisboa, Portugal} \thanks{%
CMAF, Complexo Interdisciplinar, Universidade de Lisboa, Av. Gama Pinto, 2 -
1649-003 Lisboa, Portugal, e-mail: vilela@cii.fc.ul.pt}}
\maketitle

\begin{abstract}
Many signals in Nature, technology and experiment have a multi-component
structure. By spectral decomposition and projection on the eigenvectors of a
family of unitary operators, a robust method is developed to decompose a
signals in its components. Different signal traits may be emphasized by
different choices of the unitary family. The method is illustrated in
simulated data and on data obtained from plasma reflectometry experiments in
the tore Supra.
\end{abstract}

\section{Introduction}

Most natural and man-made signals are nonstationary and may be thought of as
having a multicomponent structure. Bat echolocation, whale sounds, radar,
sonar and many others are examples of this kind of signals. The notion of
nonstationarity is easy to define. However, the concept of signal component
is not so clearly defined. Because time and frequency descriptions are
standard methods of signal analysis, many authors have attempted to base the
characterization of signal components on the analysis of the time-frequency
plane. There is a large class of time-frequency signal representations
(TFR). An important set of such TFR's is Cohen's class\cite{Cohen1},
obtained by convolutions with the Wigner distribution 
\begin{equation}
C_{\Phi }\left( t,f\right) =\int \int W\left( u,v\right) \Phi \left(
t-u,f-v\right) dudv  \label{In1}
\end{equation}
$W\left( u,v\right) $ being the Wigner distribution 
\begin{equation}
W\left( t,f\right) =\int s\left( t+\frac{\tau }{2}\right) s^{*}\left( t-%
\frac{\tau }{2}\right) e^{-i2\pi \tau f}d\tau  \label{In2}
\end{equation}
Once one particular TFR of the signal is constructed, the search for
components may be done by looking for amplitude concentrations in the
time-frequency plane. This is the methodology that has been followed by most
authors\cite{Cohen2} \cite{Khadra} \cite{Choi} \cite{Fineberg} \cite{Jones} 
\cite{Francos} \cite{Flandrin} \cite{Boashash} \cite{Rong} \cite{Wang}. The
notions of instantaneous frequency and instantaneous bandwidth play an
important role in these studies.

An important drawback from the use of TFR's is the fact that they may have
negative terms, cross terms or lack the correct marginal properties in time
and frequency. Even if, by the choice of a clever kernel or a smoothing or
filtering operation, the TFR's are apparently free from these problems,
there is no guarantee that they are free from artifacts that might lead to
unwarranted inferences about the signal properties. This is a consequence of
the basic fact that time $\left( t\right) $ and frequency $\left( \omega =i%
\frac{d}{dt}\right) $, being associated to a pair of noncommuting operators,
there can never be a joint probability distribution in the time-frequency
plane.

Our approach to \textit{component separation} starts from the insight that
the notion of \textit{component} depends as much on the observer as on the
observed object. That is, when we speak about a component of a signal we are
in fact referring to a particular feature of the signal that we want to
emphasize. For example, if time and frequency are the features that interest
us, it might indeed be the salient features in the time-frequency plane to
be identified as components. However, if it is frequency and fractality
(scale) that interests us, the notion of component and the nature of the
decomposition would be completely different.

In general, the features that interest us correspond to incompatible notions
(that is, to noncommuting operators). Therefore to look for robust
characterizations in a joint feature plane is an hopeless task because the
noncommutativity of the operators precludes the existence of joint
probabilities. Instead, in our approach, we consider spectral decompositions
using the eigenvectors of linear combinations of the operators. The sum of
the squares of the signal projections on these eigenvectors has the same
norm as the signal, thus providing an exact probabilistic interpretation.
Important operator linear combinations are the time-frequency 
\begin{equation}
B^{(S)}\left( \mu ,\nu \right) =\mu t+\nu \omega =\mu t+\nu i\frac{d}{dt}
\label{In3}
\end{equation}
the frequency-scale 
\begin{equation}
B_{1}^{(A)}\left( \mu ,\nu \right) =\mu \omega +\nu D=\mu \omega +\nu \frac{1%
}{2}\left( t\omega +\omega t\right)  \label{In4}
\end{equation}
and the time-scale 
\begin{equation}
B_{2}^{(A)}\left( \mu ,\nu \right) =\mu t+\nu D  \label{In5}
\end{equation}
Then, a quadratic positive signal transform is defined by 
\begin{equation}
M_{s}^{B}\left( X,\mu ,\nu \right) =\int s^{*}\left( t\right) \delta \left(
B\left( \mu ,\nu \right) -X\right) s\left( t\right) dt  \label{In6}
\end{equation}
called a $B-$tomogram which, for a normalized signal 
\begin{equation}
\int \left| s\left( t\right) \right| ^{2}dt=1  \label{In7}
\end{equation}
is also normalized 
\begin{equation}
\int M_{s}^{B}\left( X,\mu ,\nu \right) dX=1  \label{In8}
\end{equation}
For each $\left( \mu ,\nu \right) $ pair, the tomograms $M_{s}^{B}\left(
X,\mu ,\nu \right) $ provide a probability distribution on the variable $X$,
corresponding to a linear combination of the chosen operators (time and
frequency, frequency and scale or time and scale). Therefore, by exploring
the family of operators for all pairs $\left( \mu ,\nu \right) $ one obtains
a robust (probability) description of the signal at all intermediate
operator combinations.

Using the (symmetric) operators $B\left( \mu ,\nu \right) $ and their
corresponding unitary exponentiations 
\begin{equation}
U\left( \mu ,\nu \right) =\exp \left( iB\left( \mu ,\nu \right) \right)
\label{In9}
\end{equation}
a unified description of all currently known integral transforms has been
obtained\cite{JPA}. Explicit expressions for the tomograms in the three
cases (\ref{In3}-\ref{In5}) may be found in \cite{PLA}.

Of particular interest for the component analysis in this paper is the
time-frequency operator $B^{(S)}\left( \mu ,\nu \right) $ for which 
\begin{equation}
M_{s}\left( x,\mu ,\nu \right) =\frac{1}{2\pi |\nu |}\left| \int s(t)\exp
\left( \frac{i\mu }{2\nu }\,t^{2}-\frac{ix}{\nu }\,t\right) \,dt\right| ^{2}
\label{In10}
\end{equation}
called the \textit{symplectic tomogram}. The tomogram is an homogeneous
function 
\begin{equation}
M_{s}\left( \frac{x}{p},\frac{\mu }{p},\frac{\nu }{p}\right) =\left|
p\right| M_{s}\left( x,\mu ,\nu \right)  \label{In11}
\end{equation}

For the particular case $\mu =\cos \theta ,\nu =\sin \theta $, the
symplectic tomogram coincides with the Radon transform\cite{Gelfand},
already used for signal analysis by several authors\cite{Wood1} \cite{Wood2} 
\cite{Barbarossa} in a different context.

Once a tomogram for $B=\mu O_{1}+\nu O_{2}$ is constructed, what one obtains
in the $\left( X,\left( \mu ,\nu \right) \right) $ (hyper-) plane is an
image of the probability flow from the $O_{1}-$description of the signal to
the $O_{2}-$description, through all the intermediate steps of the linear
combination. In contrast with the time-frequency representations we need not
worry about cross-terms or artifacts, because of the exact probability
interpretation of the tomogram. Then, we may define as a \textit{component
of the signal} any distinct feature (ridge, peak, etc.) of the probability
distribution in the $\left( X,\left( \mu ,\nu \right) \right) $ (hyper-)
plane. It is clear that the notion of component is contingent on the choice
of the pair $\left( O_{1},O_{2}\right) $.

In Sect.2 we analyze in detail the time-frequency tomogram, the choice of a
complete orthogonal basis of eigenvectors of $B^{S}\left( \mu ,\nu \right) $
for the projection of the signal and how the component identification may be
carried out by spectral decomposition into subsets of this basis. In Sect.3
a few examples of component decomposition of noisy signals are worked out,
which show the effectiveness of the method. Finally, in Sect.4 the method is
applied to experimental data obtained in the reflectometry analysis of
plasma density. In the Appendixes we collect a few results, which are useful
for the practical calculation of the symplectic tomograms.

\section{Tomograms and Signal Analysis}

Following the ideas described in the introduction, a probability family of
distributions, $M_{s}(x,\theta )$, is defined from a (general) complex
signal $s(t)$, $t\in [0,T]$ by 
\begin{equation}
M_{s}(x,\theta )=\left| \int \ s(t)\Psi _{x}^{\theta ,T}(t)\,dt\right|
^{2}=\left| <s,\Psi _{x}^{\theta ,T}>\right| ^{2}  \label{eq:2.1}
\end{equation}
with 
\begin{equation}
\Psi _{x}^{\theta ,T}(t)=\frac{1}{\sqrt{T}}\exp \left( \frac{-i\cos \theta }{%
2\sin \theta }\,t^{2}+\frac{ix}{\sin \theta }\,t\right)  \label{eq:2.2}
\end{equation}
This is a particular case of Eq.(\ref{In10}) for $\mu =\cos \theta ,\nu
=\sin \theta $. Here $\theta $ is a parameter that interpolates between the
time and the frequency operators, thus running from $0$ to $\pi /2$ whereas $%
x$ is allowed to be any real number. Notice that the $\Psi _{x}^{\theta ,T}$%
's are generalized eigenfunctions for any spectral value $x$ of the operator 
$U({\theta })$. Therefore $M_{s}(x,\theta )$ is a (positive) probability
distribution as a function of $x$ for each ${\theta }$. From an abstract
point of view, since for different $\theta $'s the $U({\theta })$ (see Eq.(%
\ref{In9})) are unitarily equivalent operators, all the tomograms share the
same information. However, from a practical point of view the situation is
somehow different. In fact when $\theta $ changes from $0$ to $\pi /2$ the
information on the time localization of the signal will gradually
concentrate on large $x$ values which are unattainable because of sampling
limitations. On the other hand and by opposite reasons, close to ${\theta =0}
$ the frequency information is lost. Therefore we search for intermediate
values of ${\theta }$ where a good compromise may be found. For such
intermediate values, as we shall see in several examples, it is possible to
pull apart different components of the signal that take into account both
time and frequency information. The reason why this is the case will be
clear by looking at the properties of~(\ref{eq:2.2}).\newline
First we select a subset $x_{n}$ in such a way that the corresponding family 
$\left\{ \Psi _{x_{n}}^{\theta ,T}(t)\right\} $ is orthogonal and
normalized, 
\begin{equation}
<\Psi _{x_{m}}^{\theta ,T},\Psi _{x_{n}}^{\theta ,T}>=\delta _{m,n}
\label{eq:2.3}
\end{equation}
This is possible by taking the sequence 
\begin{equation}
x_{n}=x_{0}+\frac{2n\pi }{T}\sin \theta  \label{eq:2.4}
\end{equation}
where $x_{0}$ is freely chosen (in general we take $x_{0}=0$ but it is
possible to make other choices, depending on what is more suitable for the
signal under study).

A glance at the shape of the functions~(\ref{eq:2.2}) shows that the nodes
(the zero crossings) $t_{n}$ of the real (resp. imaginary) part of $\Psi
_{x_{n}}^{\theta ,T}$ are the solutions of 
\begin{equation}
\frac{\cos \theta }{2\sin \theta }\,t_{n}^{2}-\frac{x}{\sin \theta }%
\,t_{n}=2\pi n\qquad (\mathnormal{resp. }2\pi n+\pi /2)  \label{eq:2.5}
\end{equation}

It is clear that $\left| t_{n+1}-t_{n}\right| $ scales as $\sqrt{n}$ and
that, for fixed $\theta $, the oscillation length at a given $t$ decreases
when $\left| x\right| $ increases. As a result, the projection of the signal
on the $\left\{ \Psi _{x_{n}}^{\theta ,T}(t)\right\} $ will locally explore
different scales. On the other hand, changing $\theta $ will modify the
first term of (\ref{eq:2.5}) in such a way that the local time scale is
larger when $\theta $ becomes larger in agreement with the uncertainty
principle.

We then consider the projections of the signal $s(t)$ 
\begin{equation}
c_{x_{n}}^{\theta }(s)=<s,\Psi _{x_{n}}^{\theta ,T}>  \label{eq:2.6}
\end{equation}
which in the following are used for signal processing purposes. In
particular a natural choice for denoising consists in eliminating the $%
c_{x_{n}}^{\theta }(s)$ such that 
\begin{equation}
\left| c_{x_{n}}^{\theta }(s)\right| ^{2}\leq \epsilon  \label{eq:2.7}
\end{equation}
for some chosen threshold $\epsilon $, the remainder being used to
reconstruct a denoised signal. In this case a proper choice of $\theta $ is
an important issue in the method.

In the present work we mainly explore the spectral decomposition of the
signal to perform a multi-component analysis. This is done by selecting
subsets $\mathcal{F}_{k}$ of the $x_{n}$ and reconstructing partial signals (%
$k$-components) by restricting the sum to 
\begin{equation}
s_{k}(t)=\sum_{n\in \mathcal{F}_{k}}c_{x_{n}}^{\theta }(s)\Psi
_{x_{n}}^{\theta ,T}(t)  \label{eq:2.8}
\end{equation}
for each $k$.

Eq.(\ref{eq:2.8}) builds the signal components as spectral projections of $s$%
. As we shall see, by an appropriate choice of $\theta $, it is possible to
use this technique to disentangle the different components of a signal. This
work is mainly devoted to the analysis of reflectometry signals in plasma
physics for which the method seems well adapted.

Notice that the inverse of $\left| t_{n+1}^{\theta }-t_{n}^{\theta }\right| $
plays the role of a quasi-instantaneous frequency defined in a $\theta $%
-like scale. This is a piecewise constant function but, as seen from (\ref%
{eq:2.2}), it grows approximately linearly in time with slope $\tan
^{-1}(\theta )$. We used such time scales to control the quality of the
sampling.

\section{Examples: Simulated data}

In this section we discuss the general method presented in the previous
section in two particular simulated signals. The first example shows how the
method is able to disentangle a signal with different time as well as
frequency components. In the second example a signal with time-varying
frequency is analyzed.

\subsection{First Example}

\begin{figure}[htb]
\begin{center}
\includegraphics[height=6cm,width=9cm]{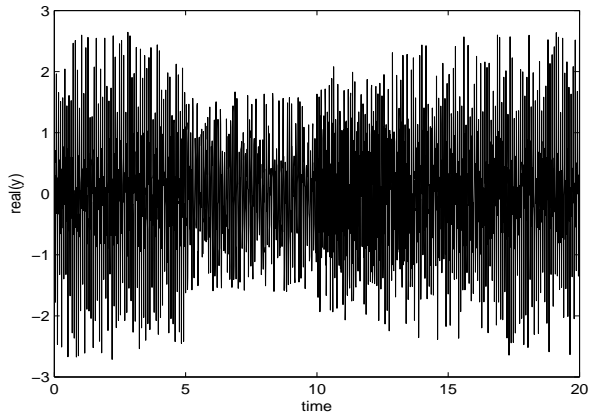}
\end{center}
\caption{Temporal representation of the simulated data $\mathcal{R}\left[
y(t)\right] $ defined by equation(\protect\ref{eq:3.1.1}).}
\label{Fi:sin-y}
\end{figure}

Let us consider a signal $y(t)$, of duration $T=20$ $s$, that is the sum of
three sinusoidal complex signals $y_{k}$, $k=1,2,3$, plus a noise component $%
b$ : 
\begin{equation}
y(t)=y_{1}(t)+y_{2}(t)+y_{3}(t)+b(t)  \label{eq:3.1.1}
\end{equation}
where 
\begin{eqnarray*}
y_{1}\left( t\right) &=&\exp \left( i25t\right) ,t\in \left[ 0,20\right] \\
y_{2}\left( t\right) &=&\exp \left( i75t\right) ,t\in \left[ 0,5\right] \\
y_{3}\left( t\right) &=&\exp \left( i75t\right) ,t\in \left[ 10,20\right]
\end{eqnarray*}
The Signal to Noise Ratio, $SNR_{y,b},$ is about $10$ $dB$, the SNR being
defined by 
\begin{equation}
SNR(y,b)=10\log _{10}\frac{P_{y}}{P_{b}}  \label{eq:3.1.2}
\end{equation}
with $P_{y}=\frac{1}{T}\int_{0}^{T}\left| y(t)\right| ^{2}dt$ and $P_{b}=%
\frac{1}{T}\int_{0}^{T}\left| b(t)\right| ^{2}dt$. The real part of the
simulated data, $\mathcal{R}\left[ y(t)\right] $, is shown in the Fig.\ref{Fi:sin-y}. 
In order to test the robustness of the projection protocol we first compare
the original signal $y(t)$ with a reconstructed signal $\tilde{y}(t)$ given
by 
\begin{equation}
\tilde{y}(t)=\sum_{x_{n}=-175}^{175}c_{x_{n}}^{\theta }(y)\Psi
_{x_{n}}^{\theta ,T}(t)  \label{eq:3.1.3}
\end{equation}
The quadratic error $E(y,\tilde{y})$, between the original and the
reconstructed signal is less than $-27dB$. The quadratic error is defined as
: 
\begin{equation}
E(y,\tilde{y})=10\log _{10}\frac{P_{y-\tilde{y}}}{P_{y}}  \label{eq:3.1.4}
\end{equation}
The quadratic error grows to $-22dB$ if the reconstruction is limited to the
range used for the component analysis below (i.e. $45\leq x_{n}\leq {155}$).

\begin{figure}[tbh]
\begin{center}
\includegraphics[height=5cm,width=9cm]{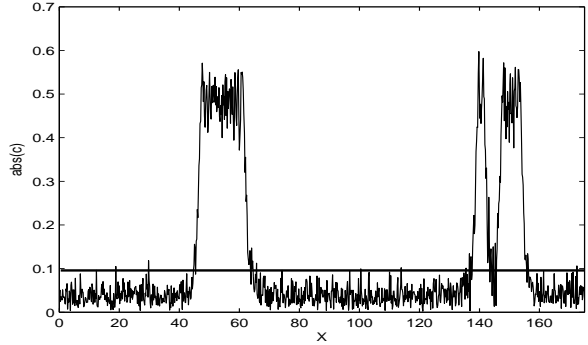}
\end{center}
\caption{$c_{x_{n}}^{\protect\theta }(y)$ spectrum of the simulated data $%
y(t)$ for $\protect\theta =\frac{\protect\pi }{5}$ and $0\leq {x_{n}}\leq 175
$}
\label{Fi:sin-cn-pi_5-partial}
\end{figure}

The value $\theta =\frac{\pi }{5}$ is chosen by direct inspection of the
tomogram of the signal $y(t)$. In this case one sees three well
differentiated spectral components (Fig.\ref{Fi:sin-cn-pi_5-partial}).
Clearly this is not the unique possible (even logical) choice. From a
practical point of view, either some preliminary information is available on
the structure of the signal, or we may try different choices of $\theta $
knowing that the incertitude on the time support will increase with $\theta $
whereas the quasi-local frequency incertitude will decrease.

In this example, we performed the factorization of $y(t)$ in three
components $\tilde{y_{1}}(t)$, $\tilde{y_{2}}(t)$, and $\tilde{y_{3}}(t)$
defined respectively by the equations (\ref{eq:3.1.5}), (\ref{eq:3.1.6}) and
(\ref{eq:3.1.7}) . Using different values of $\theta $, the quadratic errors 
$E(y,\tilde{y})$, $E(y_{1},\tilde{y_{1}})$, $E(y_{2},\tilde{y_{2}})$ and $%
E(y_{3},\tilde{y_{3}})$ are computed (Eq. \ref{eq:3.1.4}). We summarize the
corresponding data in the following table:

\begin{center}
\begin{tabular}{|c|c|c|c|c|c|}
\hline
$\theta$ & $\pi$/8 & $\pi$/5 & 3$\pi$/10 & 4$\pi$/5 & $\pi$/2 \\ \hline
$E(y_{1},\tilde{y}_{1})$ & -14.5dB & -17.5dB & -18.5dB & -17.5dB & -12.5dB
\\ \hline
$E(y_{2},\tilde{y}_{2})$ & -10.5dB & -12.5dB & -9dB & -7dB & -0.5dB \\ \hline
$E(y_{3},\tilde{y}_{3})$ & -14.5dB & -14dB & -13.5dB & -7dB & -4dB \\ \hline
$E(y,\tilde{y})$ & -26.5dB & -27dB & -30dB & -30dB & -28dB \\ \hline
\end{tabular}
\end{center}

In cases where no a priori information is available, different choices of $%
\theta $ may provide meaningful information about the signal structure. In
this example, by looking at the data presented on the table above, the
choice of $\theta =\frac{\pi }{5}$ to carry out the factorization seems to
give the best performance. Then we simply apply an energy threshold $%
\epsilon =0.1$, which is about $15\%$ of the energy level of the signal, to
decompose the signal $y$ in three components (Fig.\ref{Fi:sin-cn-pi_5-partial}).

The first component, $\tilde{y_{1}}(t)$ corresponds to the spectral range $%
45\leq x_{n}\leq 65$: 
\begin{equation}
\tilde{y_{1}}(t)=\sum_{x_{n}=45}^{65}c_{x_{n}}^{\theta }(y)\Psi
_{x_{n}}^{\theta ,T}(t)  \label{eq:3.1.5}
\end{equation}

The second component, $\tilde{y_{2}}(t)$ corresponds to the spectral range $%
135\leq x_{n}\leq 145$: 
\begin{equation}
\tilde{y_{2}}(t)=\sum_{x_{n}=135}^{145}c_{x_{n}}^{\theta _{0}}(y)\Psi
_{x_{n}}^{\theta _{0},T}(t)  \label{eq:3.1.6}
\end{equation}

The real part of $y_{2}(t)$ and $\tilde{y_{2}}(t)$ are presented Fig.\ref%
{Fi:sin-y2-pi_5}.

\begin{figure}[htb]
\begin{center}
\includegraphics[height=5cm,width=9cm]{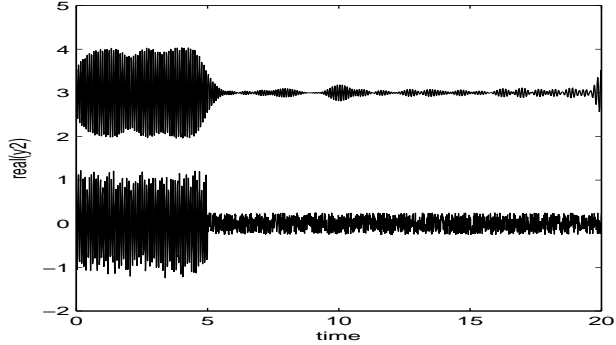}
\end{center}
\caption{$\mathcal{R}\left[y_2(t)\right]$ and $\mathcal{R}\left[\tilde{y_2}%
(t)\right]$. For visual purposes, the mean value of $\mathcal{R}%
\left[\tilde{y_2}(t)\right]$ is shifted to 3.}
\label{Fi:sin-y2-pi_5}
\end{figure}

The last component, $\tilde{y_{3}}(t)$ corresponds to the spectral range $%
145\leq x_{n}\leq 155$: 
\begin{equation}
\tilde{y_{3}}(t)=\sum_{x_{n}=145}^{155}c_{x_{n}}^{\theta }(y)\Psi
_{x_{n}}^{\theta ,T}(t)  \label{eq:3.1.7}
\end{equation}

\begin{figure}[htb]
\begin{center}
\includegraphics[height=5cm,width=9cm]{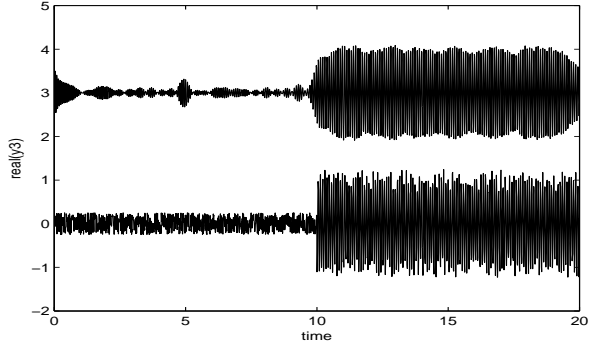}
\end{center}
\caption{Representation of $\mathcal{R}\left[y_3(t)\right]$ and $\mathcal{R}%
\left[\tilde{y_3}(t)\right]$.\textit{\ For visual purposes, the mean value
of $\mathcal{R}\left[\tilde{y_3}(t)\right]$ is shifted to 3.} }
\label{Fi:sin-y3-pi_5}
\end{figure}

Fig.\ref{Fi:sin-y3-pi_5} gives a representation of both $\mathcal{R}\left[
y_{3}(t)\right] $ and $\mathcal{R}\left[ \tilde{y_{3}}(t)\right] $. 

The quadratic errors $E(y_{1},\tilde{y_{1}})$, $E(y_{2},\tilde{y_{2}})$ and $%
E(y_{3},\tilde{y_{3}})$ can be read from the table above. They are,
respectively, $-17.5dB$, $-12.5dB$ and $-14dB$.

\begin{figure}[htb]
\begin{center}
\includegraphics[height=5cm,width=9cm]{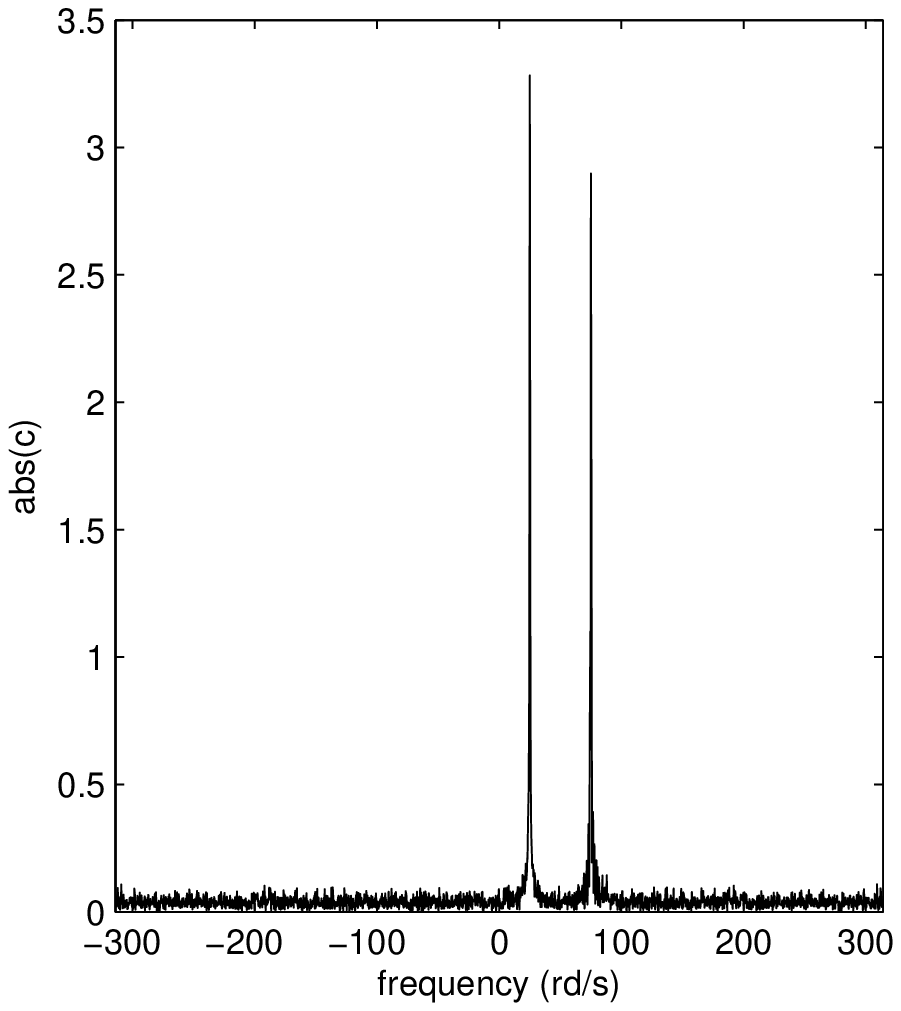}      
\end{center}
\caption{Projection $c^{\theta_0}_{x_n}$ of the signal for $\theta_0=\frac{\pi}{2}$ }
\label{Fi:sin-fourier}
\end{figure}

For comparison, the projection of the simulated data $y(t)$ in the frequency domain ($\theta_0=\frac{\pi}{2}$), presented on figure \ref{Fi:sin-fourier}, show that the factorization in three components is not possible : only two components can be extract from this projection. At the frequency $x_n=25 rd/s$, the component will be equal to $\tilde{y_1}(t)$. At the frequency  $x_n=75 rd/s$, it will be impossible to set apart $y_2(t)$ and $y_3(t)$ and the component will be equal to  $\tilde{y_2}(t)+\tilde{y_3}(t)$.

%


\subsection{Second Example}

Here we analyze the decomposition into elementary components of another
signal which aims to mimic, in a simplified way, the case of an incident
plus a reflected wave delayed in time and with an acquired time-dependent
change in phase. In this case the simulated signal $y(t)$ is the sum of an
``incident'' chirp $y_{0}(t)$ and a ``deformed reflected'' chirp $y_{R}(t)$.
White noise is added to the signal. The incident chirp is: 
\begin{equation}
y_{0}(t)=e^{i\Phi _{0}(t)}
\end{equation}
with $\Phi _{0}(t)=a_{0}t^{2}+b_{0}t$.

The ``instantaneous frequency'' of $y_{0}(t)$ sweeps linearly from $75$ $rd/s
$ to $50$ $rd/s$ during $20s$. Its phase derivative is linearly dependent on time :$\frac{d}{dt}\Phi
_{0}(t)=2a_{0}t+b_{0}$.

The ``reflected'' signal $y_{R}(t)$ is delayed by $t_R=3s$ from the incident
one and continuously sweeps from $75rd/s$ to $50rd/s$ : 
\begin{equation}
y_{R}(t)=e^{i\Phi _{R}(t)}
\end{equation}
where $\Phi _{R}(t)=a_{R}(t-t_R)^{2}+b_{R}(t-t_R)+10(t-t_R)^{\frac{3}{2}}$.
In this case the phase derivative $\frac{d}{dt}\Phi _{R}(t)$ is not a linear
function. This signal is zero during the first $3s$ seconds and ends up at $%
t=23s$.

The simulated signal is defined by : 
\begin{equation}
y(t)=y_0(t)+y_R(t)+ b(t)  \label{eq:3.2.1}
\end{equation}
The Signal to Noise Ratio, $SNR(y,b)$, is equal to $15dB$. The real $%
\mathcal{R}\left[y(t)\right]$ part of this signal is shown in Fig\ref%
{Fi:chirp-y}. 
\begin{figure}[htb]
\begin{center}
\includegraphics[height=6cm,width=9cm]{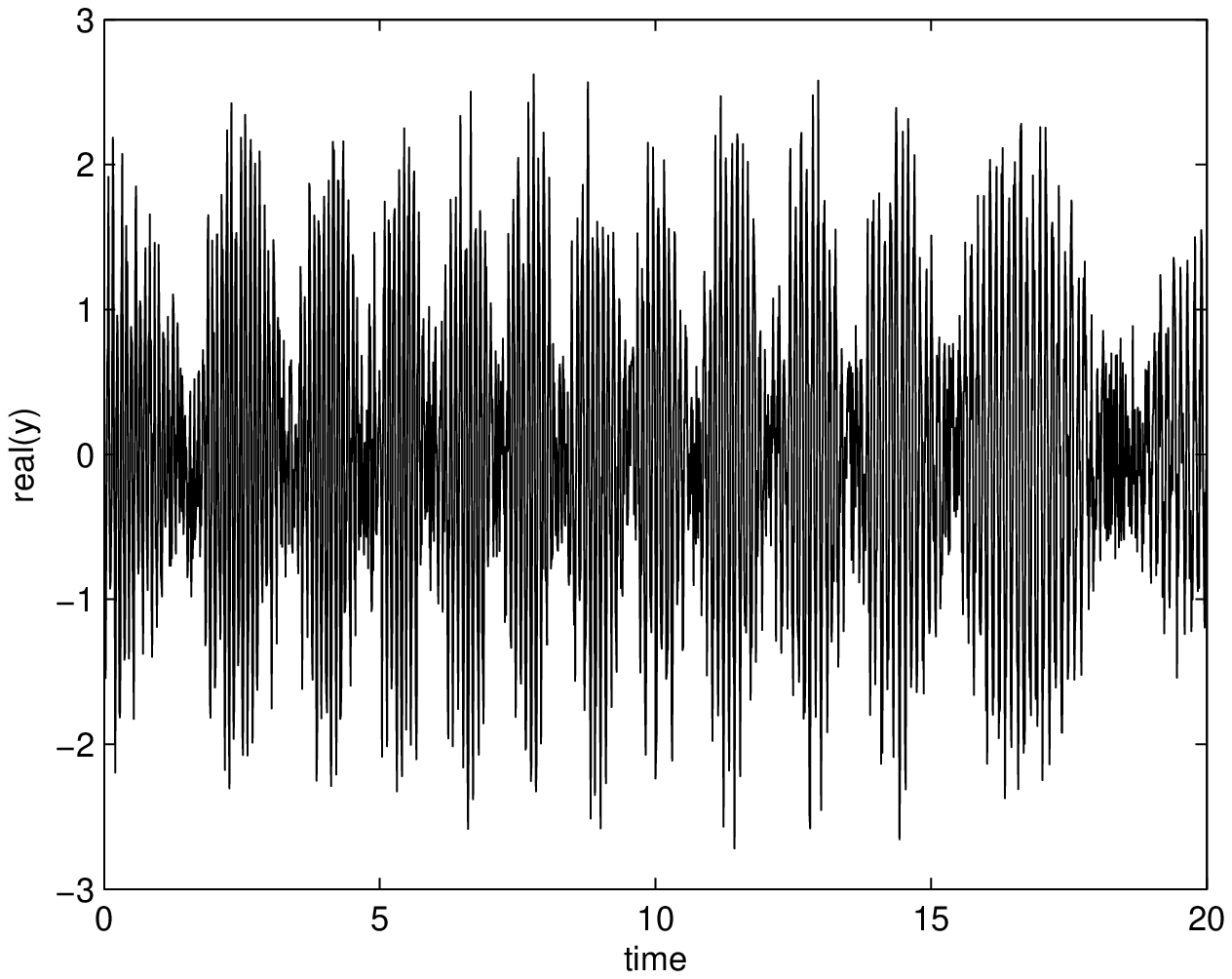}
\end{center}
\caption{Temporal representation of the simulated data $\mathcal{R}\left[y(t)%
\right]$ defined by equation(\protect\ref{eq:3.2.1}).}
\label{Fi:chirp-y}
\end{figure}

\begin{figure}[htb]
\begin{center}
\includegraphics[height=6cm,width=9cm]{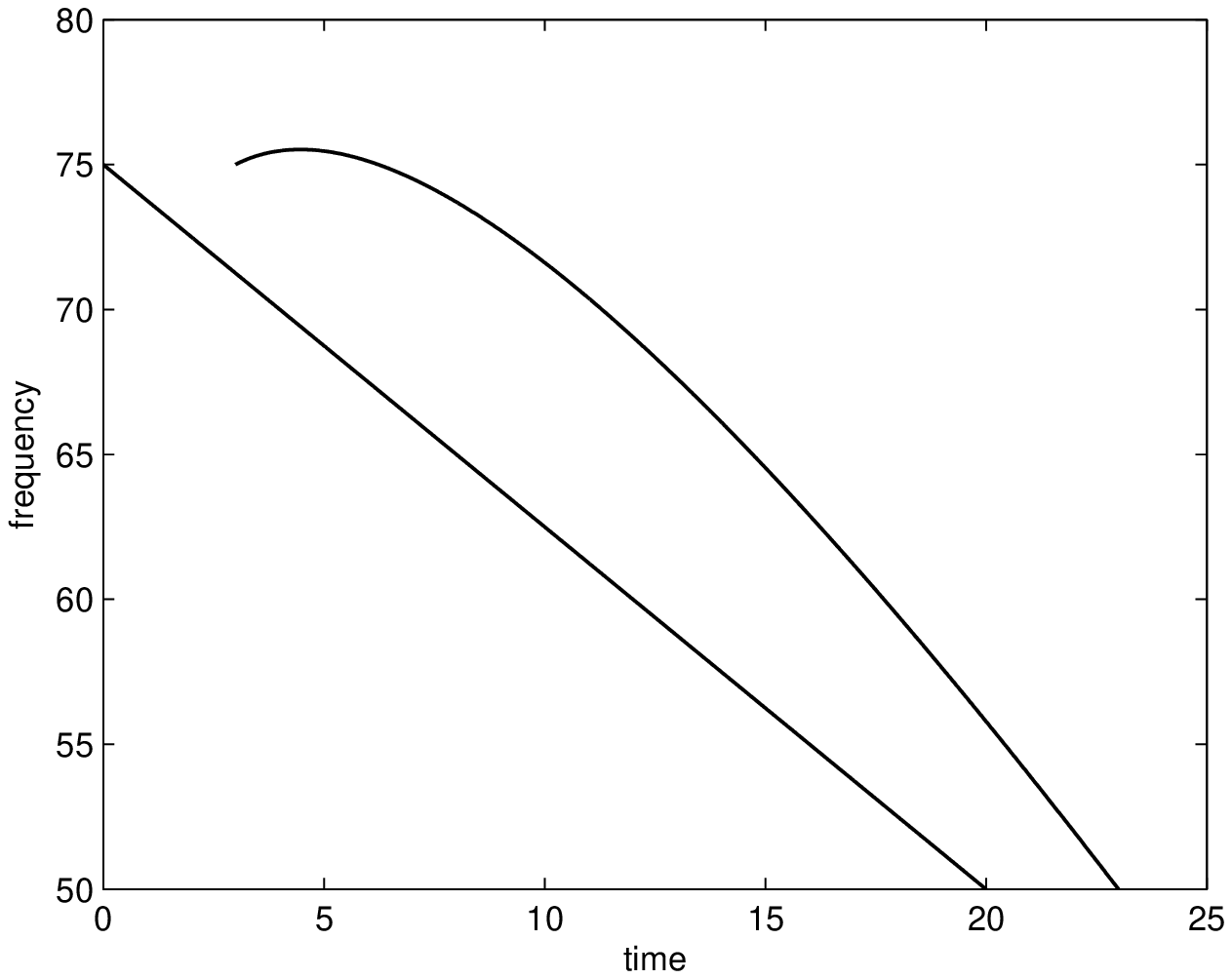}
\end{center}
\caption{Representation of $\frac{d}{dt} \Phi_0 (t)$ and $\frac{d}{dt}
\Phi_R (t)$ as a function of time. }
\label{Fi:chirp-finst}
\end{figure}

Fig.\ref{Fi:chirp-finst} shows $\frac{d}{dt}\Phi _{0}(t)$ and $\frac{d}{dt}%
\Phi _{R}(t)$ as a function of time. Notice that, except for the three first
seconds, the spectrum of the signals $y_{0}(t)$ and $y_{R}(t)$ is almost the
same. 

\begin{figure}[htb]
\begin{center}
\includegraphics[height=6cm,width=9cm]{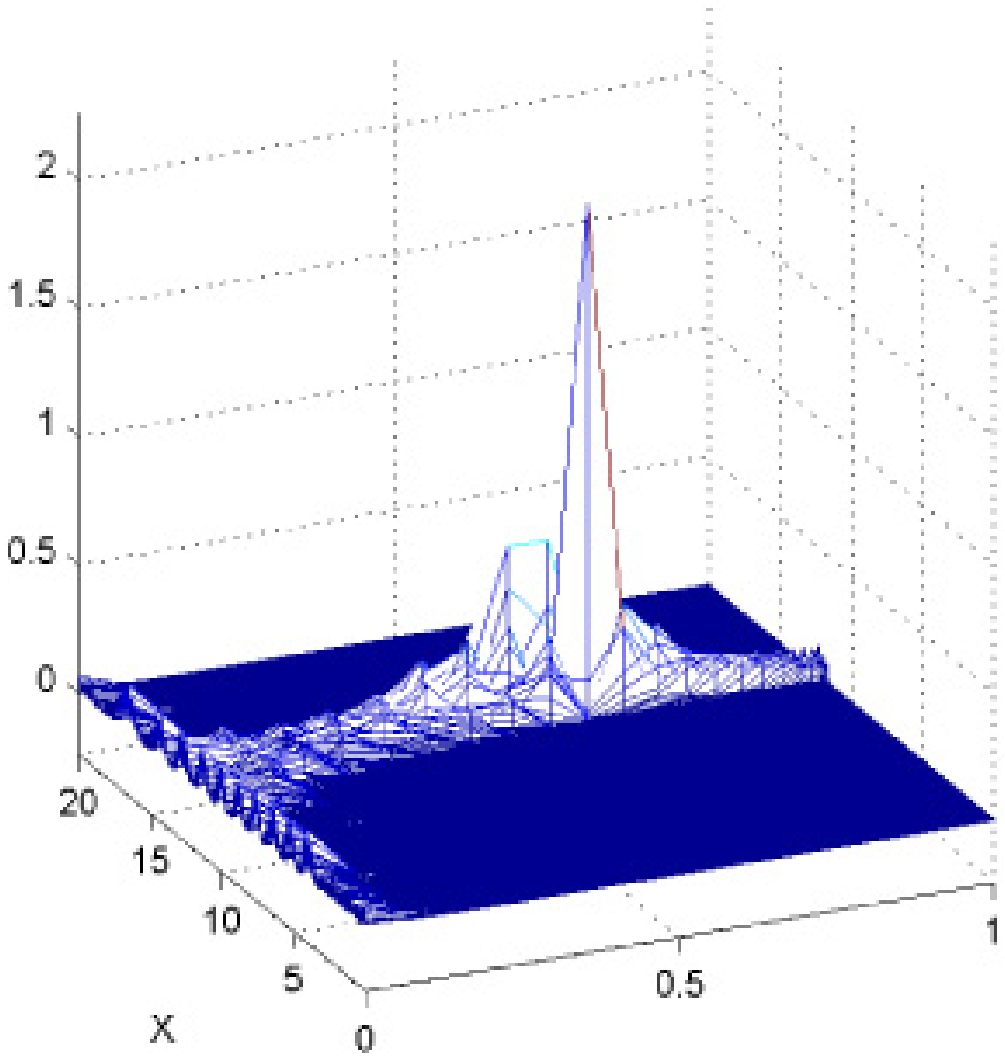}
\end{center}
\caption{Tomogram of the simulated data defined by Eq.(\protect\ref{eq:3.2.1})}
\label{Fi:tomo-chirp}
\end{figure}

The tomogram of the first $20s$ of $y(t)$, $M_{y}(\theta ,x)=\left| <y,\Psi
_{x}^{\theta ,T}>\right| ^{2}$ has a maximum for $\sin (\theta )\approx 0.6$
(Fig.\ref{Fi:tomo-chirp}) corresponding to the ``incident'' part of the
signal that mainly projects in the unique $\Psi _{x}^{\theta ,T}$ that
matches $\Phi _{0}(t)$. Then, we take the value of $\theta =\pi /5$ to carry
out the separation of $y(t)$ in its components. 

The corresponding spectrum $c_{x_{n}}^{\theta }(y)$ is shown in Fig.\ref%
{Fi:chirp-cn-pi_5}. Based on this spectrum we decompose the signal in two
spectral components. 
\begin{figure}[htb]
\begin{center}
\includegraphics[height=6cm,width=9cm]{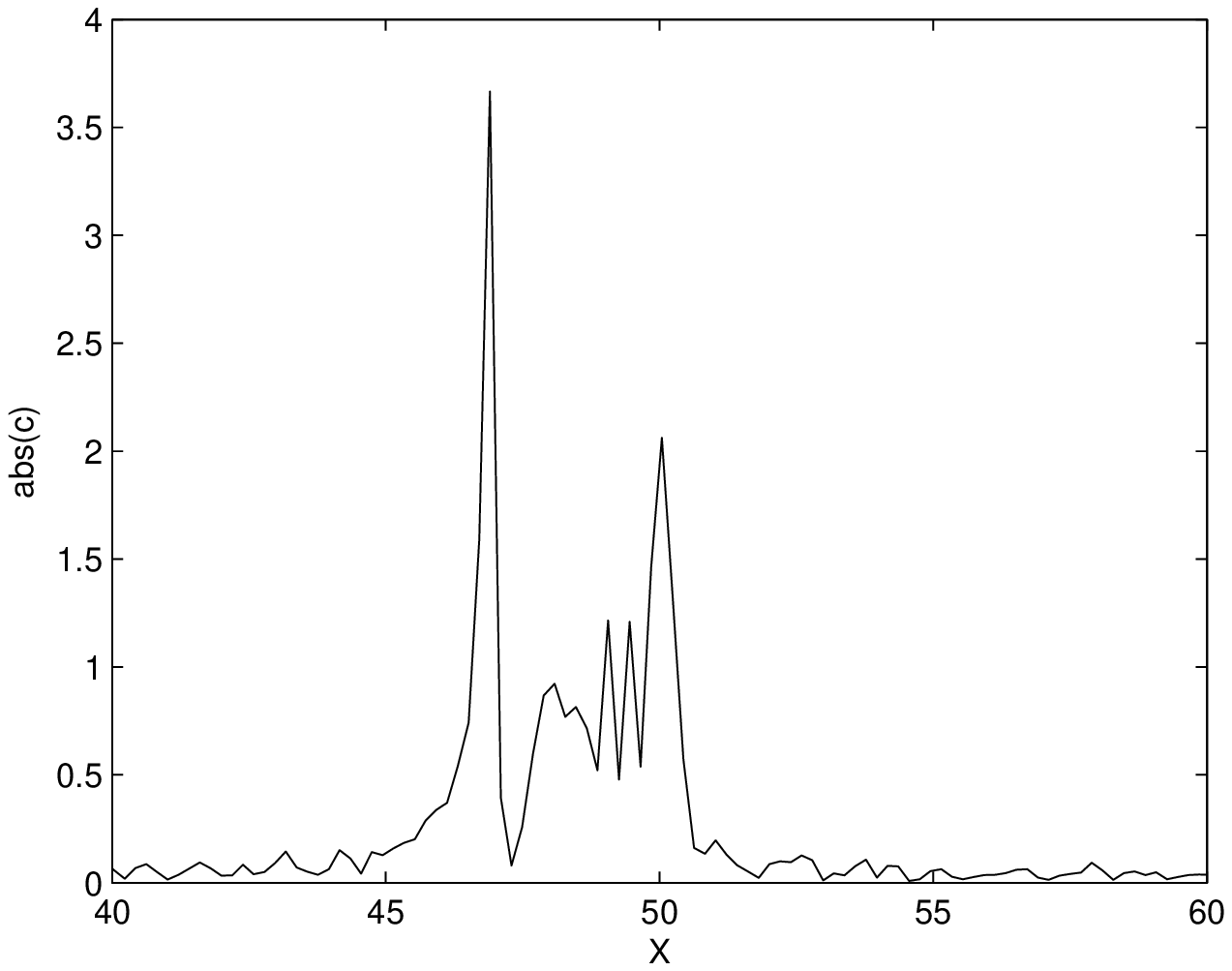}
\end{center}
\caption{Spectrum of the signal for $\protect\theta=\protect\pi/5$}
\label{Fi:chirp-cn-pi_5}
\end{figure}

From the first component we reconstruct the \textquotedblleft
incident\textquotedblright\ chirp $y_{0}(t)$ by: 
\begin{equation}
\tilde{y_{0}}(t)=\sum_{x_{n}=45}^{47.25}c_{x_{n}}^{\theta _{0}}(y)\Psi
_{x_{n}}^{\theta _{0},T}(t)  \label{3.2.2}
\end{equation}%
The quadratic error, between $\tilde{y_{0}}(t)$ and $y_{0}(t)$, $E(y_{0},%
\tilde{y_{0}})$, is $-9.5dB$.

From the second spectral component we reconstruct the \textquotedblleft
reflected\textquotedblright\ chirp given by: 
\begin{equation}
\tilde{y_{R}}(t)=\sum_{x_{n}=47.5}^{50.5}c_{x_{n}}^{\theta }(y)\Psi
_{x_{n}}^{\theta ,T}(t)  \label{3.2.3}
\end{equation}%
In this case the quadratic error $E(y_{R},\tilde{y_{R}})$ is $-10dB$. This
may be compared with a quadratic error $E(y,\tilde{y})$ of $-29dB$ for the
total signal reconstructed from the spectral projection corresponding to $%
45<x_{n}<50.5$.

We have tested the method with different delays that encode for the distance
of the two \textquotedblleft frequencies\textquotedblright . The quality of
the disentanglement deteriorates when the delay decreases. But it can still
be done for a delay as short as $1s$. Therefore we conclude that the method
is quite robust.

\section{An application to reflectometry data. Component analysis}

We now use signals coming from reflectometry measurement in plasma physics,
to show the ability of the tomogram methods to separate different components
of the signal to which it is then possible to assign a clear physical
meaning. The reflectometry diagnostic is widely used to determine the
electronic density profile in a tokamak. The principle, based upon a radar
technique \cite{Hugenholtz}, is to measure the phase of a probing wave
reflected by the plasma cut--off layer at a given density, where the
refractive index goes to zero. The determination of the density profile can
be achieved by continuously sweeping the frequency of the probing wave.

Different techniques are used to measure the density profile on fusion
plasmas \cite{Laviron} (phase difference, ultrashort pulses, continuous
sweep, ...). A broadband reflectometer operating in the frequency range
50--75 GHz (V band)~\cite{Clairet1},~\cite{Clairet2} and 75--110 GHz (W
band)~\cite{Clairet3} has been developed on Tore Supra to measure the
electron density profiles at the edge.

The sweep frequency reflectometry system of Tore Supra launches a probing
wave on the extraordinary mode polarization (X mode) in the V band (50--75
GHz). The emitting and receiving antennas are located at about 1.20 m from
the plasma edge outside the vacuum vessel. The reflectometry system operates
in burst mode, i.e. the sweeps are performed repeatedly every $25\mu s$. The
duration of one sweep, $E_{0}(t)=A_{0}e^{i\phi (t)}$, is $20\mu s$ and 5000
chirps are send during one measurement. During the $20\mu s$ measurement
time, the frequency of the probing wave is continuously sweeping, from 50
GHz to 75 GHz (V band).

The heterodyne reflectometers, with I/Q detection, provide a good Signal to
Noise Ratio, up to $40dB$. For each sweep, the reflected chirp $E_{R}(t)$ is
mixed with the incident sweep $E_{0}(t)$ and only the interference term is
recorded as in-phase and $90^{\circ }$ phase shifted signals sampled at $%
T_{e}=10^{-8}s$ 
\begin{equation*}
x_{1}(t)=A_{0}A_{R}(t)\cos (\varphi (t)) 
\end{equation*}
\begin{equation*}
x_{2}(t)=A_{0}A_{R}(t)\sin (\varphi (t)) 
\end{equation*}

\begin{figure}[htb]
\begin{center}
\includegraphics[height=6cm,width=9cm]{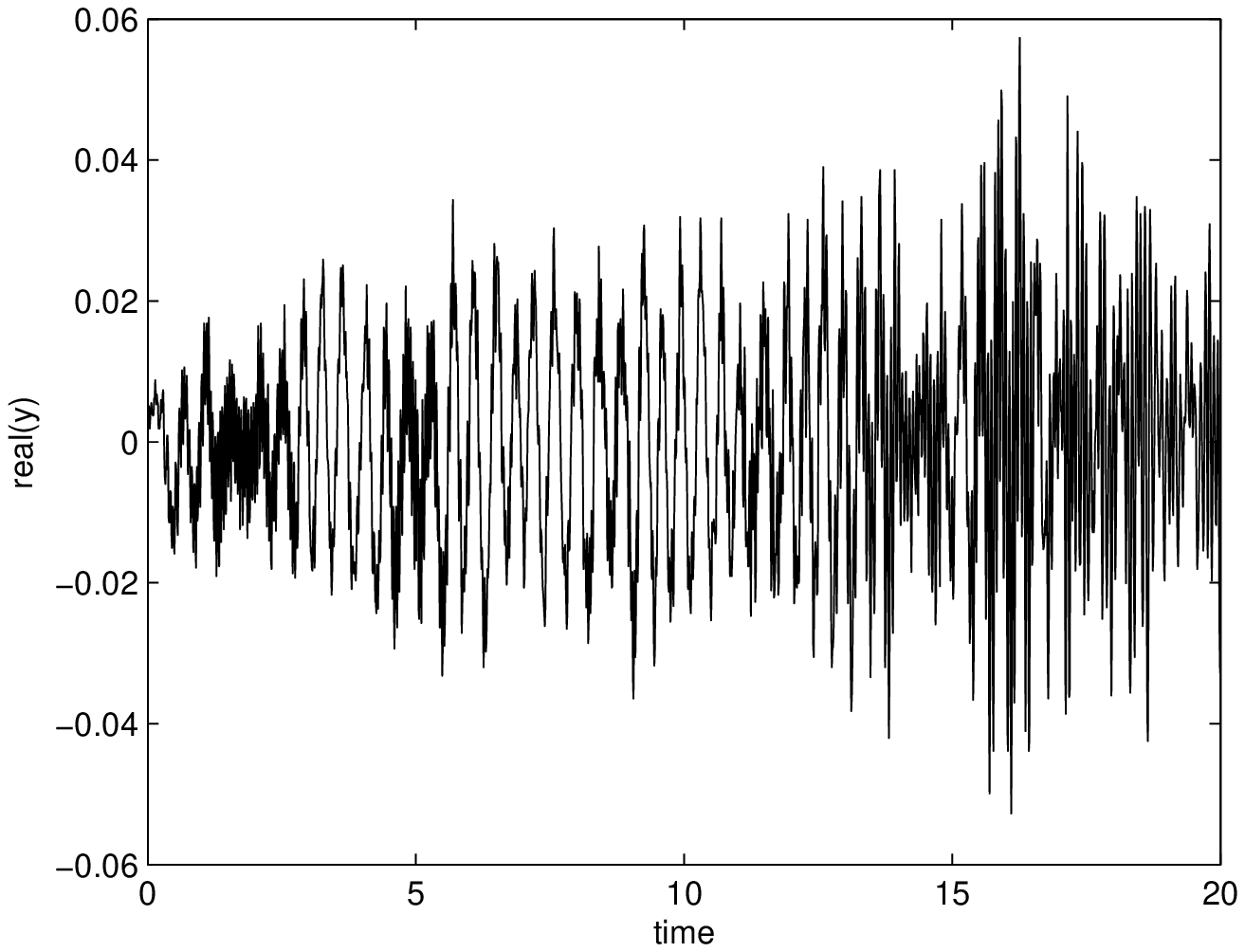}
\end{center}
\caption{Time representation of the reflectometry signal. }
\label{Fi:reflec-y}
\end{figure}

For each sweep, the phase $\varphi (t)$ of the reflected signal is
represented by 
\begin{equation}
y(t)=x_1(t)+ ix_2(t)=A(t)e^{i\varphi (t)}  \label{eq:4.1}
\end{equation}
The amplitude of this signal $A(t)=A_{0}A_{R}(t)$ is low frequency. The real
part of one such signal $y(t)$ is shown in Fig.\ref{Fi:reflec-y}.

\begin{figure}[htb]
\begin{center}
\includegraphics[height=5cm,width=9cm]{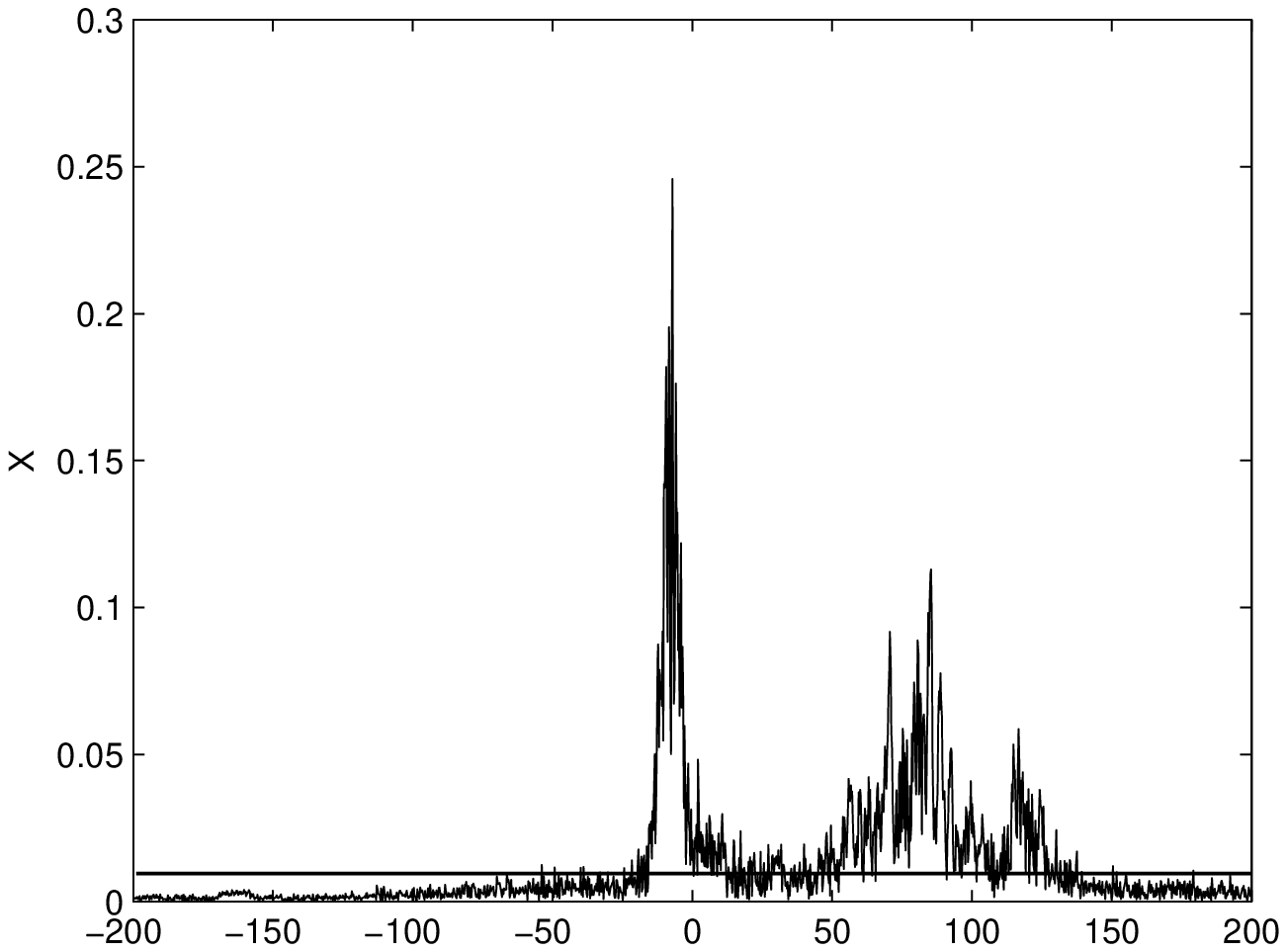}
\end{center}
\caption{Spectrum $c^{\protect\theta}_{x_n}$ of the reflectometry signal $%
y(t)$ for $\protect\theta=\frac{3 \protect\pi}{10}$ }
\label{Fi:reflec-cn-3pi_10}
\end{figure}

The tomogram $M_{y}(x,\theta )$ of the signal is shown in Fig.\ref%
{Fi:tomo-reflec} where it is possible to see that it carries three main
different components. 
\begin{figure}[htb]
\begin{center}
\includegraphics[height=6cm,width=9cm]{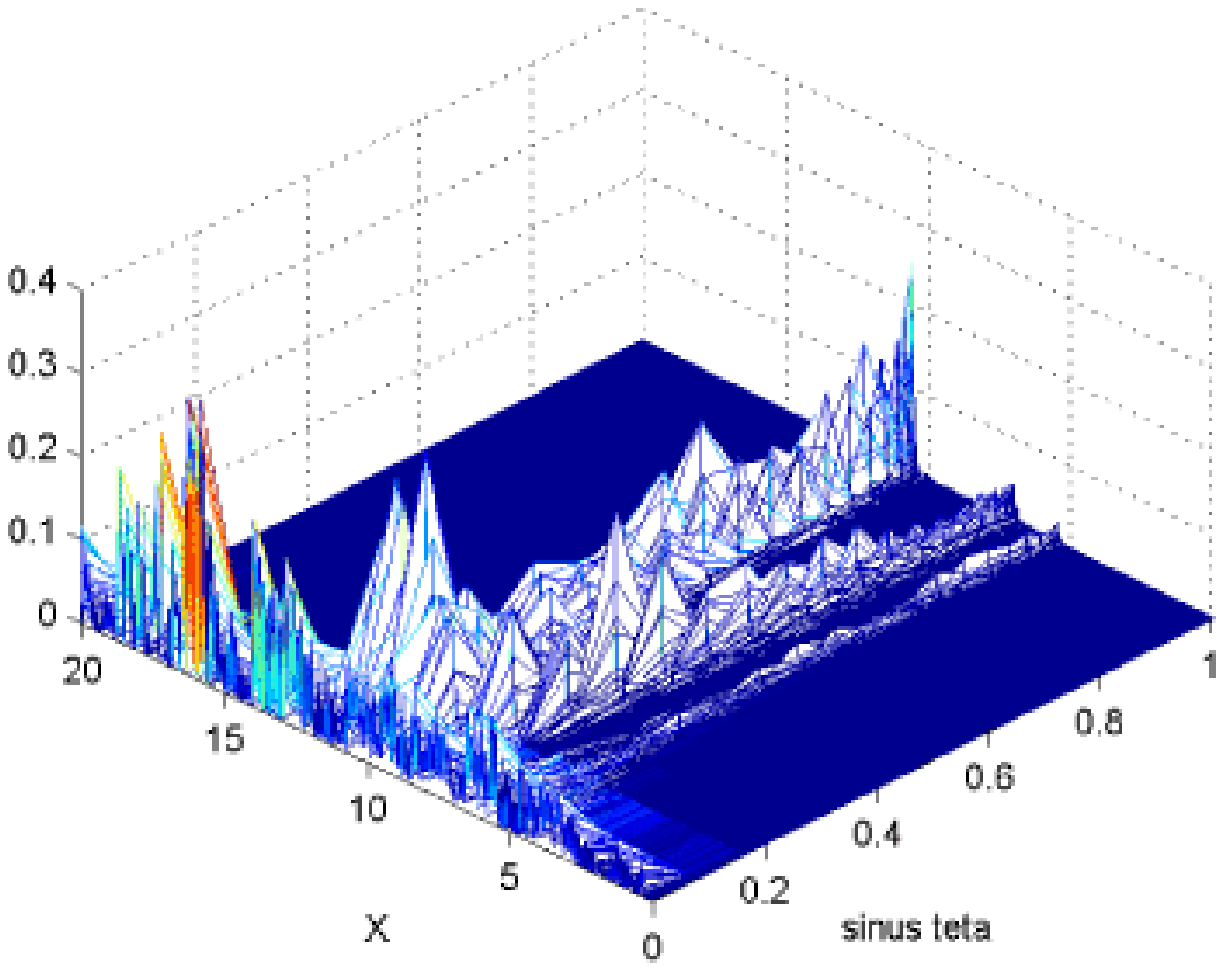}
\end{center}
\caption{Tomogram of the reflectometry signal. }
\label{Fi:tomo-reflec}
\end{figure}
The choice of $\theta =\frac{3\pi }{10}$ to perform the factorization of the
signal was done by simple inspection of this tomogram. The spectrum $%
c_{x_{n}}^{\theta }(y)$ of the reflectometry signal for $\theta =\frac{3\pi 
}{10}$ is shown in Fig.\ref{Fi:reflec-cn-3pi_10}.

When reconstructing $\tilde{y}(t)$ by: 
\begin{equation}
\tilde{y}(t)=\sum_{x_{n}=-200}^{200}c_{x_{n}}^{\theta }(y)\Psi
_{x_{n}}^{\theta ,T}(t)  \label{eq:4.2}
\end{equation}
the quadratic error $E(y,\tilde{y})$, between the original and the
reconstructed signals is $-25dB$.

\textbf{Factorization of the reflectometry signal}

\begin{figure}[htb]
\begin{center}
\includegraphics[height=5cm,width=9cm]{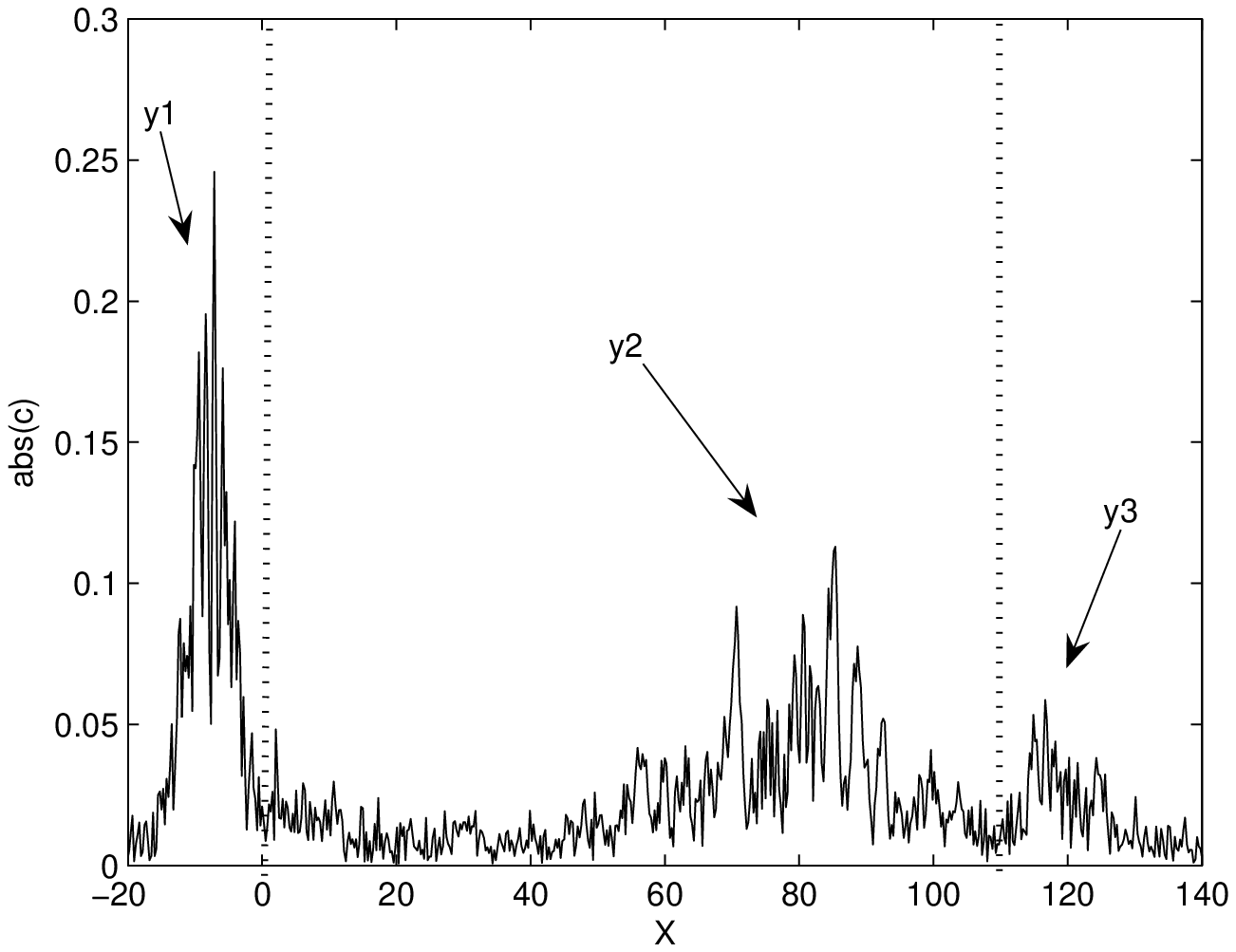}
\end{center}
\caption{Part of the spectrum $c^{\protect\theta}_{x_n}(y)$ of the
reflectometry signal used in the factorization}
\label{Fi:reflec-cn-3pi_10-partial}
\end{figure}

By taking a threshold equal to $\epsilon =0.01$ we select the spectral
components corresponding to $\left| c_{x_{n}}\right| \neq 0$ for $-20\leq
x_{n}\leq 140$ (see Fig.\ref{Fi:reflec-cn-3pi_10-partial}). The error
between the original and the selected signal is about $-18dB$. From there
the spectrum of $y(t)$ splits in three components.

\textbf{First component, the reflection on the porthole}

\begin{figure}[htb]
\begin{center}
\includegraphics[height=5cm,width=9cm]{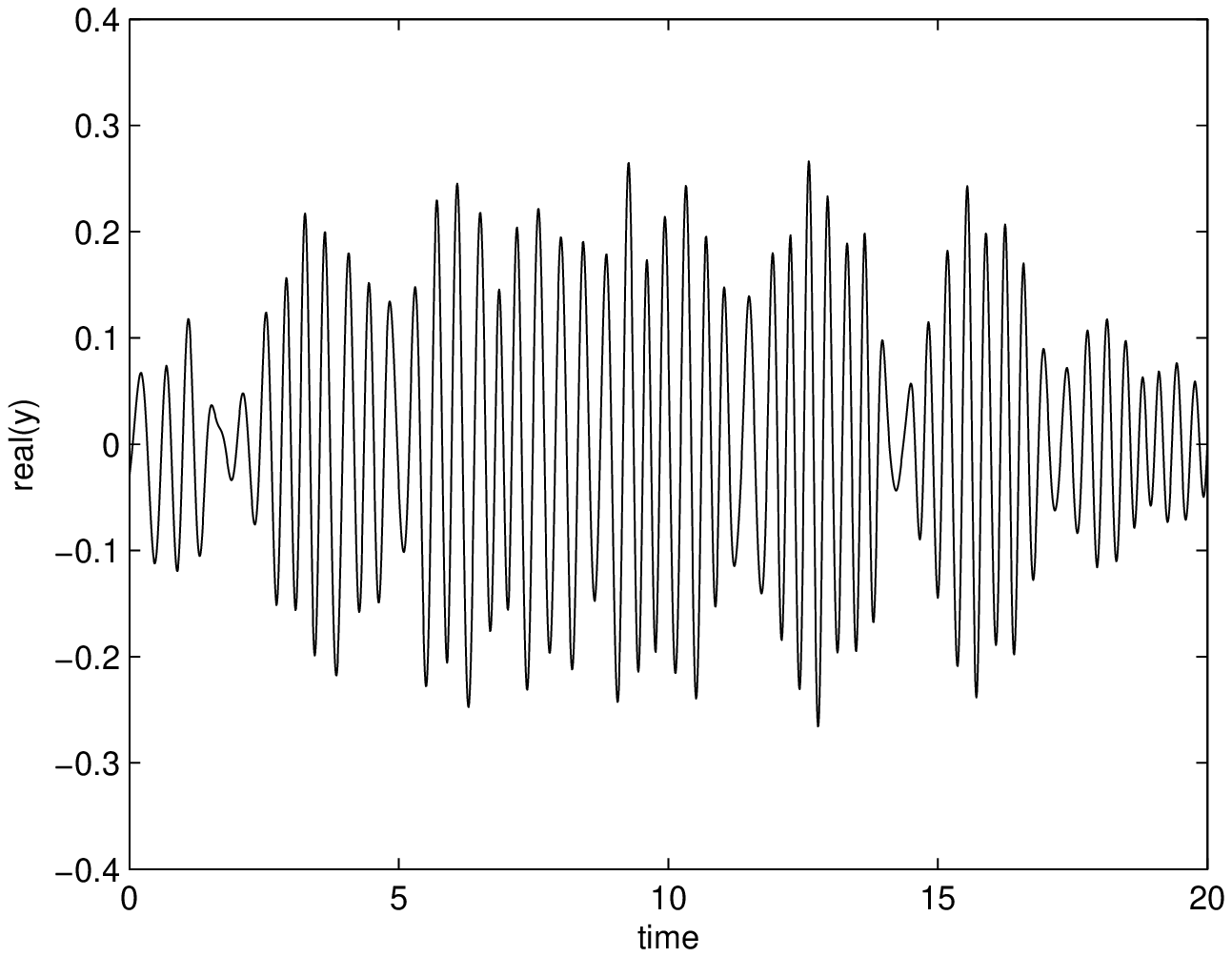}
\end{center}
\caption{First component of the reflectometry signal corresponding to the
reflection on the porthole }
\label{Fi:reflec-y1-3pi_10}
\end{figure}

The first component, $\tilde{y_{1}}(t)$ corresponds to $-20\leq x_{n}\leq 0$
and is therefore defined as: 
\begin{equation}
\tilde{y_{1}}(t)=\sum_{x_{n}=-20}^{0}c_{x_{n}}^{\theta }(y)\Psi
_{x_{n}}^{\theta }(t)  \label{eq:4.3}
\end{equation}
It is a low frequency signal corresponding to the heterodyne product of the
probe signal with the reflection on the porthole \cite{Clairet3}. It is
shown in Fig.\ref{Fi:reflec-y1-3pi_10}.

\textbf{Second component, the plasma signal}

\begin{figure}[htb]
\begin{center}
\includegraphics[height=5cm,width=9cm]{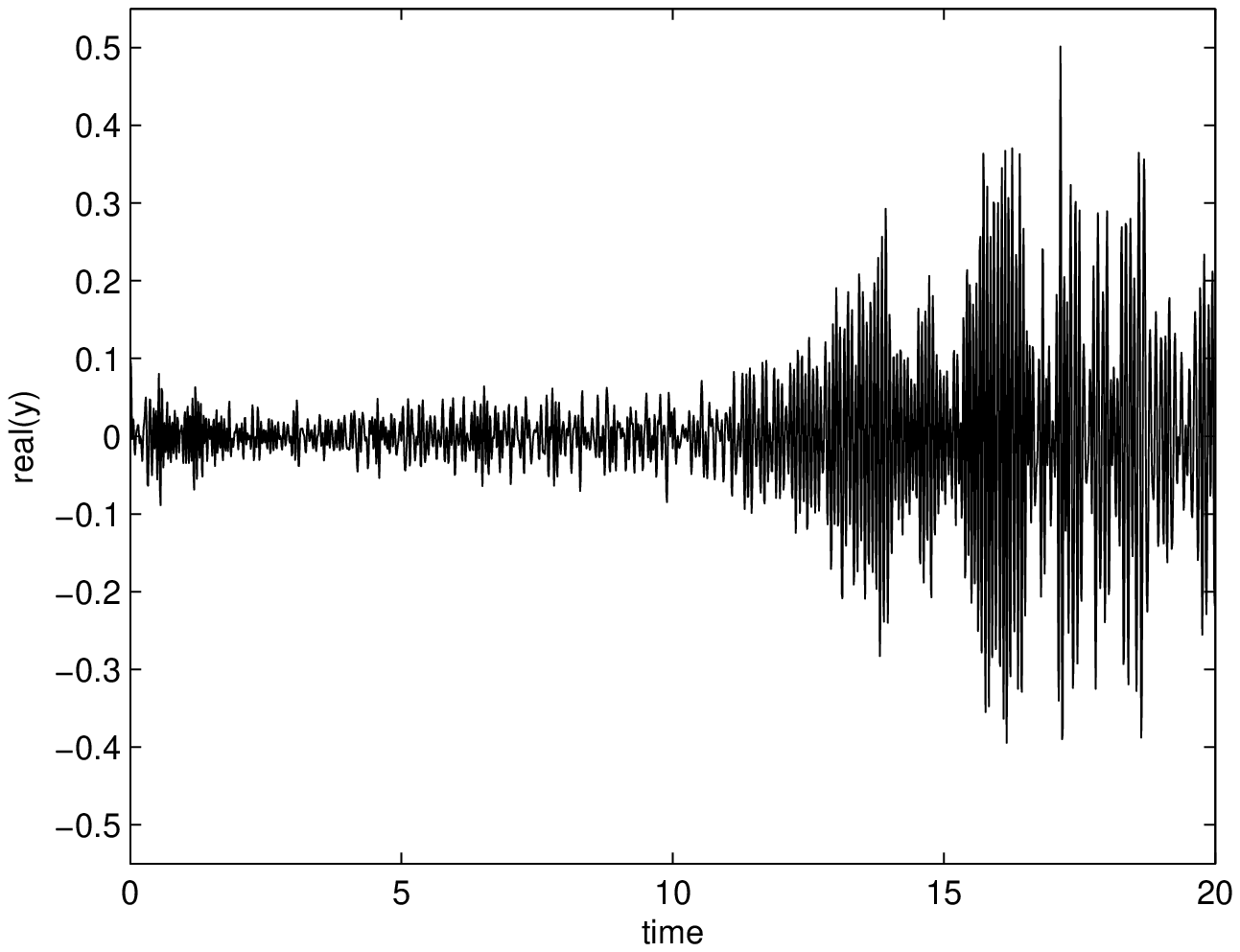}
\end{center}
\caption{Second component of the reflectometry signal, corresponding to the
reflection on the plasma}
\label{Fi:reflec-y2-3pi_10}
\end{figure}

The second component has a Fourier spectra that fits the expected behavior
corresponding to the reflection of the wave inside the plasma of the
tokamak, \cite{Clairet3}. This component, $\tilde{y_{2}}(t)$, corresponds to 
$0\leq x_{n}\leq 110$ and is therefore defined as: 
\begin{equation}
\tilde{y_{2}}(t)=\sum_{x_{n}=0}^{110}c_{x_{n}}^{\theta }(y)\Psi
_{x_{n}}^{\theta }(t)  \label{eq:4.4}
\end{equation}
It is shown in Fig.\ref{Fi:reflec-y2-3pi_10}.

\textbf{Third component, the multi-reflection component}

\begin{figure}[htb]
\begin{center}
\includegraphics[height=5cm,width=9cm]{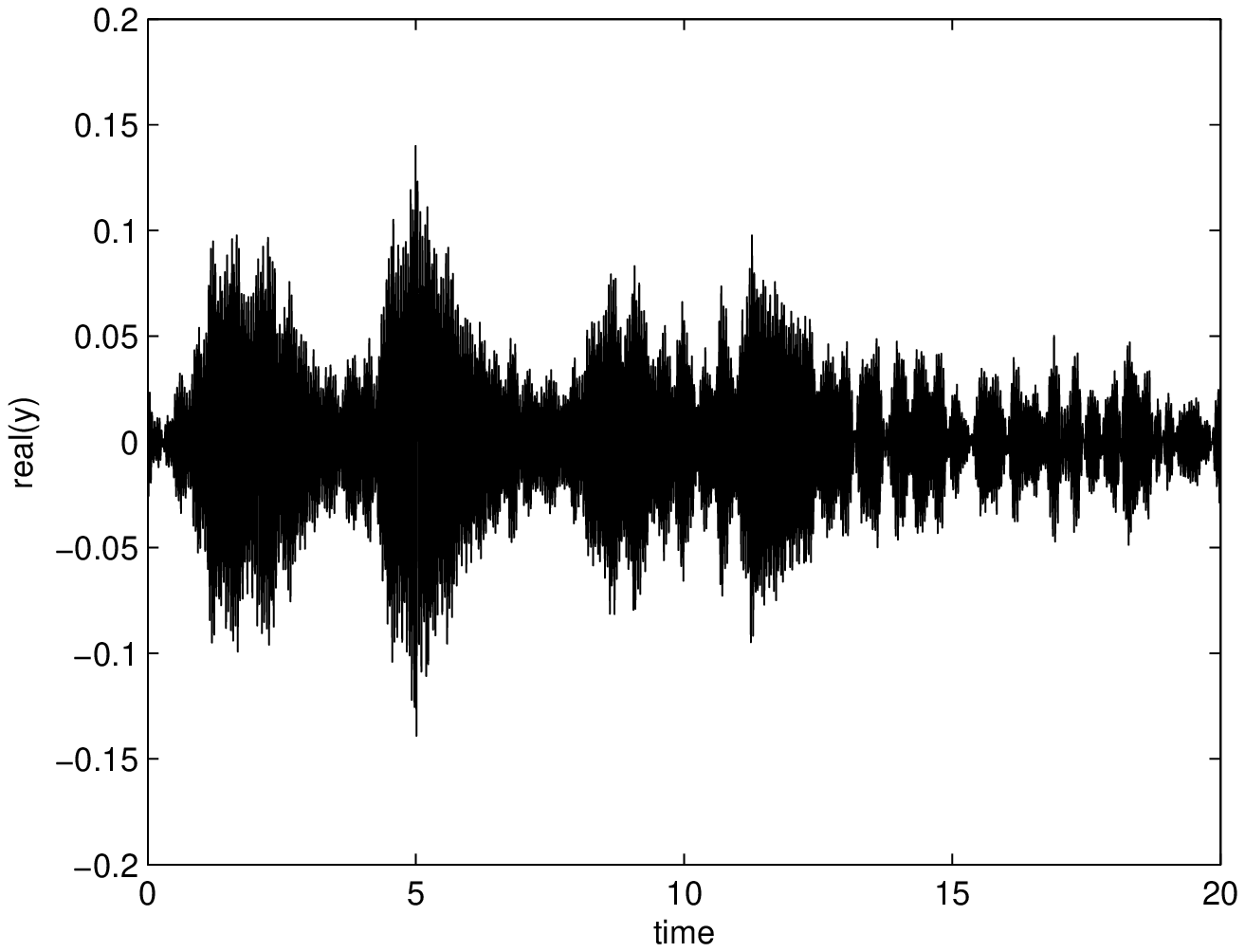}
\end{center}
\caption{Third component of the reflectometry signal, corresponding to multi
reflections on the vessel}
\label{Fi:reflec-y3-3pi_10}
\end{figure}

The last component corresponds \cite{Clairet3} to multi reflections of the
waves on the wall of the vacuum vessel. This component, $\tilde{y_{3}}(t)$,
corresponds to $110\leq x_{n}\leq 140$ and is therefore defined as: 
\begin{equation}
\tilde{y_{3}}(t)=\sum_{x_{n}=10}^{140}c_{x_{n}}^{\theta }(y)\Psi
_{x_{n}}^{\theta }(t)  \label{eq:4.5}
\end{equation}
This component is shown in Fig.\ref{Fi:reflec-y3-3pi_10}. 

We notice that by undertaking a new factorization of this third component it
seems possible to separate different successive reflections of the wave but
this would be out of the scope of this work.

\begin{figure}[tbh]
\begin{center}
\includegraphics[height=5cm,width=9cm]{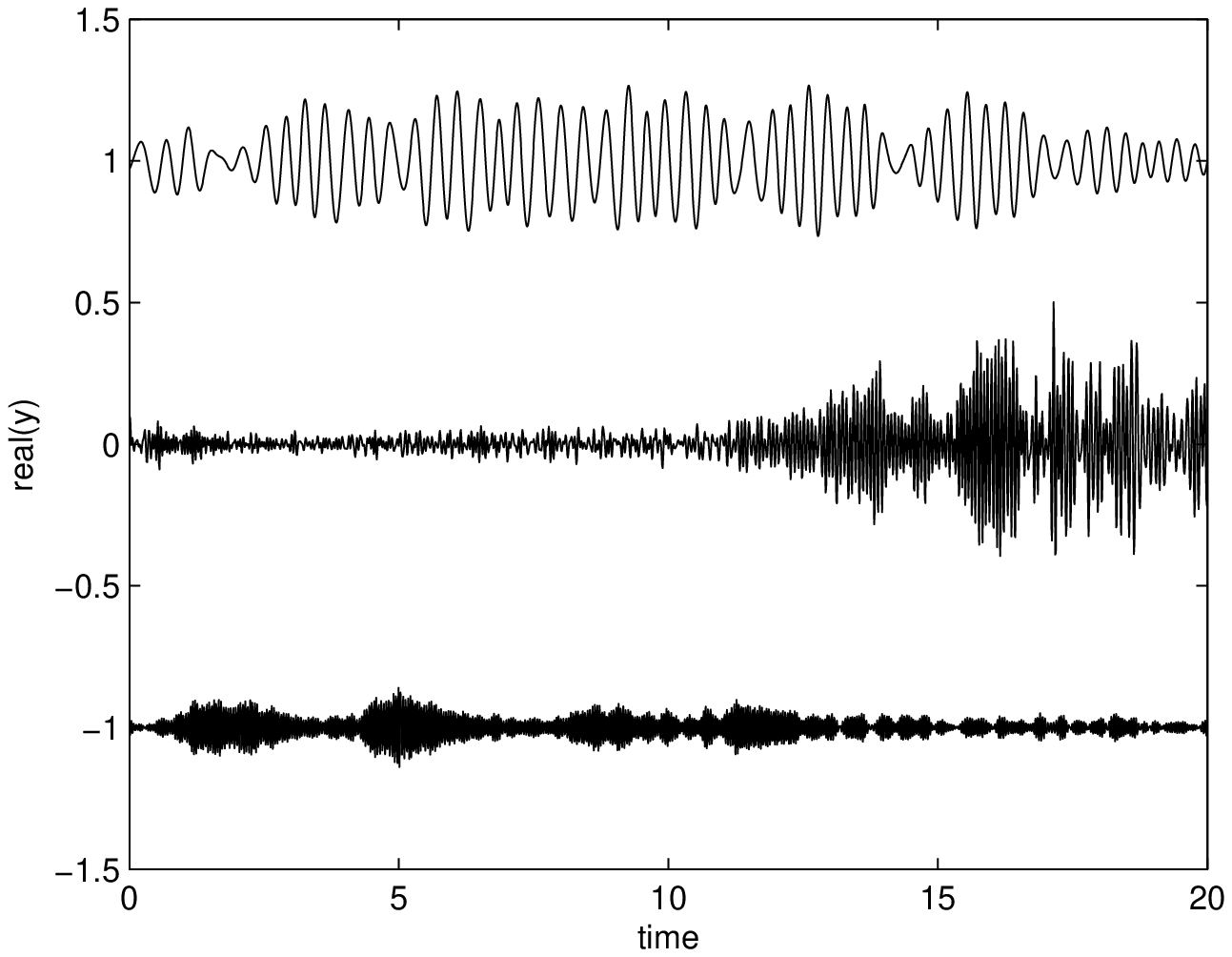}
\end{center}
\caption{The three components of the reflectometry signal. For visual purposes,
the average of $\tilde{y_{1}}(t)$ is shifted to 1 and the average of $\tilde{%
y_{3}}(t)$ to -1.}
\label{Fi:reflec-y1y2y3-3pi_10}
\end{figure}

\textit{\ }The three components of the reflectometry signal are presented
together on the same plot (Fig. \ref{Fi:reflec-y1y2y3-3pi_10}). It is
instructive to compare this factorization with the original reflectometry
signal (see Fig. \ref{Fi:reflec-y}). 

\section{Remarks and conclusions}

1) Based on a complete and probabilistically rigorous spectral analysis and
projection on the eigenvectors of a family of unitary operators, our method
seems quite robust to disentangle the relevant components of the signals.
This has been demonstrated both on simulated and on experimental
reflectometry data. In particular in this last case, a clear identification
of the physical origin of the components and its separation is readily
achieved. Such separation could not be achieved by the simple filtering
techniques. After the component separation phase, the method also provides
by truncation of some subsets of the projection coefficients a very flexible
denoising technique.

Another important conclusion from this study is the fact that by the choice
of different families of (unitary) operators and their spectral
representations, different traits and components of the signals may be
emphasized.

2) In the analysis of reflectometry data, component separation and denoising
is a required first step to obtain reliable information on the plasma
density. In particular, accuracy in these measurements is quite critical if
in addition to the average local density one also wants to have information
on plasma fluctuations and turbulence.

The location $x_{c}\left( \omega _{p}\right) $ of the reflecting (cut-off)
layer for the frequency $\omega _{p}$ is related to the group delay\cite%
{Simonet} \cite{Manso1} :
\begin{equation}
\tau _{g}=\frac{d\phi \left( \omega \right) }{d\omega }=\frac{1}{2\pi }\frac{%
d\phi }{df}  \label{4.1}
\end{equation}
and for a linear frequency sweep of the incident wave 
\begin{equation}
f\left( t\right) =f_{0}+\gamma t  \label{4.3}
\end{equation}
one obtains 
\begin{equation}
\frac{d\phi }{df}=\frac{1}{\gamma }\left. \frac{d\phi }{dt}\right| _{f}
\label{4.4a}
\end{equation}
Therefore, measurement of the plasma density hinges on an accurate
determination of the ``instantaneous frequency'' $\frac{d\phi }{dt}$.
Accuracy in the measurement of this quantity is quite critical because, the
location of the reflecting layer being obtained from an integral, errors tend to accumulate.

Several methods have been devised to obtain the group delay $\tau _{g}$ from
the reflectometry data (for a review see \cite{Manso1}). Among them,
time-frequency analysis\cite{Manso3} was widely explored. The Wigner-Ville
(WV) distribution \cite{Manso2} \cite{Bizarro}, although providing a
complete description of the signal in the time-frequency plane, raises
difficult interpretation problems due to the presence of many interference
terms that impair the readability of the distribution. For this reason the
time-frequency method that, so far, has been preferred is the spectrogram%
\cite{Manso3} \cite{Clairet} \cite{Manso4}, that is, the squared modulus of
the short-time Fourier transform 
\begin{equation*}
SP\left( t,f\right) =\left| \int_{-\infty }^{\infty }x\left( u\right)
h\left( u-t\right) e^{-i2\pi fu}du\right| ^{2} 
\end{equation*}
$h\left( u\right) $ being a peaked short-time window.

The spectrogram does not really provide the instantaneous frequency, because
that notion is not well defined anyway. All it gives is the product of the
spectra of $x\left( t\right) $ and $h\left( t\right) $. The way the
spectrogram is used to infer the local rate of phase variation $\frac{d\phi 
}{dt}$ is to identify this quantity with the maximum or the with the first
moment of the spectrogram. An additional problem comes about because
unwanted phase contributions due to plasma turbulence may have a higher
amplitude than the contributions due to the profile. Correction techniques
have been developed to compensate for this errors, based for example on
Floyd's best path algorithm. The choice of the window function is also an
important issue and, in particular, an adaptive spectrogram technique has
been developed to maximize the time-frequency concentration\cite{Manso1}.

Here, the tomographic method may also provide a needed improvement, namely
by obtaining the phase derivative directly from the coefficients $c_{n}$ of
the plasma reflection component. This, together with the selective denoising
techniques after component separation, will be described in a forthcoming
paper.

\section{Appendix A. Gauss-Hermite decomposition of the tomograms}

From the definition (\ref{In10}) of the tomogram transform one sees that the
calculation from the data near $\nu =0$ has accuracy problems because of the
fast variation of the phase in (\ref{In10}). Two techniques are used to deal
with this problem. The first one uses a projection of the time signal $%
s\left( t\right) $ on an orthogonal basis and the second uses the
homogeneity properties (\ref{In11}) and an expansion of the Fresnel tomogram
near $\nu =0$. The first technique is described in this appendix and the
second in the Appendix B.

Let $s\left( t\right) $ be a normalized signal 
\begin{equation}
\int |s(t)|^{2}dt=1.  \label{A1}
\end{equation}
Decompose the signal in Gauss-Hermite polynomials 
\begin{equation}
s(t)=\sum_{n=0}^{\infty }c_{n}(t)\psi _{n}(t),  \label{A2}
\end{equation}
with 
\begin{equation}
\psi _{n}(t)=\frac{e^{-t^{2}/2}}{\pi ^{1/4}}\,\frac{1}{\sqrt{2^{n}n!}}%
\,H_{n}(t).  \label{A3}
\end{equation}
and 
\begin{equation}
c_{n}=\int s(t)\psi _{n}(t)\,dt,  \label{A4}
\end{equation}
Then, the tomogram of the signal is 
\begin{equation}
M_{s}(X,\mu ,\nu )=M_{0}(X,\mu ,\nu )\left| \sum_{n=0}^{\infty }c_{n}\frac{1%
}{\sqrt{n!}}\left( \frac{1}{2}-\frac{1}{1-i\mu /\nu }\right)
^{n/2}H_{n}\left( \frac{b}{2\sqrt{k}}\right) \right| ^{2},  \label{A5}
\end{equation}
with 
\begin{equation}
M_{0}(X,\mu ,\nu )=\frac{1}{\sqrt{\pi \left( \mu ^{2}+\nu ^{2}\right) }}%
\,e^{-X^{2}/\left( \mu ^{2}+\nu ^{2}\right) }  \label{A6}
\end{equation}
and 
\begin{equation}
b=\frac{i\sqrt{2}X}{i\mu -\nu },\qquad k=\left( \frac{1}{2}-\frac{1}{1-i\mu
/\nu }\right) .  \label{A7}
\end{equation}

\section{Appendix B. The Fresnel tomogram}

The symplectic tomogram $M_{s}(X,\mu ,\nu )$ can be reconstructed if one
knows the (Fresnel) tomogram ~\cite{DeNicola}

\begin{equation}
M_{\mathrm{F}}(X,\nu )=M_{s}(X,1,\nu )  \label{B1}
\end{equation}
due to the homogeneity property (\ref{In11}). In fact, one has 
\begin{equation}
M_{s}(X,\mu ,\nu )=\frac{1}{|\mu |}\,M_{s}\left( \frac{X}{\mu }\,,1,\frac{%
\nu }{\mu }\right)  \label{B2}
\end{equation}
which means that, if one knows $M_{\mathrm{F}}(\widetilde{X},\widetilde{\nu }%
)$, the symplectic tomogram is obtained by replacement and a factor, 
\begin{equation}
M_{s}(X,\mu ,\nu )=\frac{1}{|\mu |}\,M_{\mathrm{F}}\left( \widetilde{X}%
\rightarrow \frac{X}{\mu }\,,\widetilde{\nu }\rightarrow \frac{\nu }{\mu }%
\right)  \label{B3}
\end{equation}

In terms of signal $s(t)$ it reads: 
\begin{eqnarray}
M_{\mathrm{F}}(X,\nu ) &=&\frac{1}{2\pi |\nu |}\left| \int
e^{i(X-y)^{2}/2\nu }s(y)\,dy\right| ^{2}  \notag \\
&=&\left| \int \frac{1}{\sqrt{2\pi |\nu |}}e^{i(X-y)^{2}/2\nu
}s(y)\,dy\right| ^{2}  \notag \\
&=&\left| \exp \left[ -i\nu \left( -\frac{1}{2}\,\frac{\partial ^{2}}{%
\partial X^{2}}\right) \right] s(X)\right| ^{2}  \label{B4}
\end{eqnarray}
Thus for small $\nu $ one has 
\begin{equation}
M_{s}(X,\mu ,\nu )\approx \frac{1}{|\mu |}\left| s\left( \frac{X}{\mu }%
\right) -\frac{i\nu }{2}\,s^{\prime \prime }\left( \frac{X}{\mu }\right)
\right| ^{2}.  \label{B5}
\end{equation}
In the Gauss-Hermite basis it is 
\begin{equation}
M_{\mathrm{F}}(X,\nu )=\frac{e^{-X^{2}/(1+\nu ^{2})}}{\sqrt{\pi (1+\nu ^{2})}%
}\left| \sum_{n=0}^{\infty }c_{n}\,\frac{1}{\sqrt{n!}}\left( \frac{1}{2}-%
\frac{1}{1-i/\nu }\right) ^{n/2}H_{n}\left( \frac{\tilde{b}}{2\sqrt{\tilde{k}%
}}\right) \right| ^{2},  \label{B6}
\end{equation}
with 
\begin{equation}
\tilde{b}=\frac{i\sqrt{2}X}{i-\nu },\qquad \tilde{k}=\left( \frac{1}{2}-%
\frac{1}{1-i/\nu }\right) .  \label{B7}
\end{equation}
As a series, the Fresnel tomogram is 
\begin{equation}
M_{F}(X,\nu )\approx \left| \sum_{k=0}^{\infty }\left( \frac{i\nu }{2}%
\right) ^{k}\,\frac{1}{k!}\,\frac{d^{2k}f(X)}{dX^{2k}}\right| ^{2},
\label{B8}
\end{equation}
leading to a symplectic tomogram 
\begin{equation}
M(X,\mu ,\nu )=\frac{1}{|\mu |}\left| \sum_{k=0}^{\infty }\left( \frac{i\nu 
}{2\mu }\right) ^{k}\,\frac{1}{k!}\frac{d^{2k}f(X/\mu )}{dX^{2k}}\right|
^{2}.  \label{B9}
\end{equation}
Since 
\begin{equation}
-\frac{1}{2}\,\frac{\partial ^{2}}{\partial t^{2}}\,s_{n}(t)+\frac{t^{2}}{2}%
\,s_{n}(t)=\left( n+\frac{1}{2}\right) s_{n}(t),\qquad n=0,1,2,\ldots ,
\label{B10}
\end{equation}
on has, for small $\nu $, the following Fresnel tomogram of $s_{n}(t)$: 
\begin{equation}
M_{n}(X,\mu =1,\nu )\approx s_{n}^{2}(X)\left[ 1+\left( n+\frac{1}{2}-\frac{%
X^{2}}{2}\right) \nu ^{2}\right]  \label{B11}
\end{equation}

\textbf{Acknowledgements}

V.I.Man'ko thanks ISEG - Lisbon (Instituto Superior de Economia e Gest$\tilde{a}$o) for
kind hospitality.

\end{document}